\documentclass[letterpaper]{article} 
\usepackage{aaai25}  
\usepackage{times}  
\usepackage{helvet}  
\usepackage{courier}  
\usepackage[hyphens]{url}  
\usepackage{graphicx} 
\usepackage{natbib}  
\usepackage{caption} 
\usepackage{algorithm}
\usepackage{algorithmic}

\usepackage{newfloat}
\usepackage{listings}
\DeclareCaptionStyle{ruled}{labelfont=normalfont,labelsep=colon,strut=off} 
\floatstyle{ruled}
\newfloat{listing}{tb}{lst}{}
\floatname{listing}{Listing}
\newcommand\href[2]{#2} 
\newcommand\hitem[3]{{\em #2} (#3)}
\newcommand\wname[1]{{\em #1}}
\newcommand\witem[2]{{\em #1} (#2)}
\newcommand\witemr[2]{{\em #2} (#1)}
\newcommand\witems[2]{{\em #1}}
\makeatletter
\renewcommand{\verbatim@font}{\small\ttfamily}
\gdef\copyright@on{}
\makeatother

\title{Class Order Disorder in Wikidata \\ and First Fixes}
\author{Peter F. Patel-Schneider\textsuperscript{\rm 1} and Ege Atacan Doğan\textsuperscript{\rm 2}}
\affiliations{\textsuperscript{\rm 1}pfpschneider@gmail.com \\
\textsuperscript{\rm 2}Faculty of Engineering and Natural Sciences, Sabancı University, Istanbul, Turkey  egeatacandogan@gmail.com}

\begin{document}

\maketitle

\begin{abstract}
Wikidata has a large ontology with classes at several orders.
The Wikidata ontology has long been known to have violations of class order
and information related to class order that appears suspect.
SPARQL queries were evaluated against Wikidata
to determine the prevalence of several kinds of violations and suspect information
and the results analyzed.
Some changes were manually made to Wikidata to remove some of these results
and the queries rerun, showing the effect of the changes.
Suggestions are provided on how the problems uncovered might be addressed,
either though better tooling or involvement of the Wikidata community.

This is an extended version of a paper submitted to AAAI 2025.
\end{abstract}

\section{Introduction}

Wikidata \cite{wikidata} is a large---110 million items as of July 2024---general-purpose, widely used ``free and open knowledge base'' that ``anyone can edit''.  
Wikidata is more than a knowledge graph as it contains a large class-based ontology
(of between 115 thousand and 17 million classes depending on how one counts) 
with instance and subclass of (generalization) links.
The Wikidata ontology covers a wide variety of domains,
with large sections devoted to
the person,
places,
life sciences \cite{wikidata-life-sciences},
medical \cite{wikidata-medical},
biological taxonomy, and genomics domains.
Wikidata is similar to other large knowledge graphs such as the Google Knowledge Graph.

Wikidata's construction by a large open community has resulted in quite a lot of incorrect information being added to it.  As Wikidata has only limited facilities for finding and fixing incorrect information and does not prioritize fixing incorrect information over adding new information these errors accumulate over time.
Some of the errors in Wikidata affect the Wikidata ontology.
Although these problems are well known 
\cite{wikidata-ontology-issues-prioritization}
insufficient effort has been made to improve the ontology so these problems 
continue and accumulate as well.
In 2024 there was an effort (\url{www.wikidata.org/wiki/Wikidata:WikiProject_Ontology/Cleaning_Task_Force})
started
to find and fix incorrect information in the Wikidata ontology involving not just this work but also work on disjointness in Wikidata and links between the Wikidata ontology and other ontologies.

One kind of problem that affects the Wikidata ontology is related to class order and related structural problems like instance and subclass loops in the ontology.
A technical definition of class order is given later in this paper but the intuition is that a class has a fixed order if all of its instances have a fixed order one lower (where non-classes, also known as individuals, have fixed order zero).
One would expect that most classes in Wikidata, like \witem{human}{Q5} has a fixed order of one, as most large domains in Wikidata are about individuals in the world (e.g., humans, animals, structures, landforms, constructed objects).
But, for example, \witem{human}{Q5} was as of June 2024 not a first-order class in Wikidata because, among other reasons, several humans were instances of themselves.
Another class order problem that affects the Wikidata ontology is classes that are stated to have more than one fixed order, which is not possible for a non-empty class.
For example, \witem{combat vehicle family}{Q100709275} was stated to be both a second-order and a third-order class.

During June 2024 queries were constructed and run against the RDF dump of Wikidata to retrieve problems related to class order
including classes that are stated to belong to multiple fixed orders,
classes that are stated to belong to a fixed order but have instances with incorrect orders,
instance and subclass loops in the ontology, 
and 
classes that are both a subclass and an instance of another class.
These queries identified many potential problems.

Sometimes the results of the queries are unequivocally errors but often the results only provide strong evidence that there is a problem and have to be further analyzed.
In a few cases where there are actual problems an update to Wikidata can be automatically constructed to fix the problem but in most cases further analysis is needed to find the correct fix. When there are many results this further analysis can be very time consuming.

For some cases where fixes can be easily identified updates were made to Wikidata and the effects of these results are shown.  For other cases suggestions are made on how they might be alleviated by either upgrading the system running Wikidata or incentivizing Wikidata communities to identify and fix problems.

\section{Wikidata}

Most of the information in Wikidata is about items, an example for an item being Douglas Adams. 
Each item in Wikidata has a unique identifier, which for Douglas Adams is Q42; a multi-lingual label, here ``Douglas Adams'' in English; and a multi-lingual description. 
Other information about items is in the form of statements---a value for
a property on an item, potentially with qualifying information\footnote{Qualifying information will be ignored in this paper as nearly all ontology statements are unqualified.}---for example,
\hitem{www.wikidata.org/wiki/Q42}{Douglas Adams}{Q42}'s 
\hitem{www.wikidata.org/wiki/Property:P22}{father}{P22}
is
\hitem{www.wikidata.org/wiki/Q14623675}{Christopher Douglas Adams}{Q14623675}.\footnote{References to most Wikidata items will include their English label and identifier.} 
The basic interface to Wikidata at \url{wikidata.org} permits searching of Wikidata and viewing and editing of the information in it.  
The information about Douglas Adams in Wikidata can be seen and edited by anyone at \url{www.wikidata.org/wiki/Q42}.

The Wikidata ontology is a class-based ontology, with classes having instances and classes being subclasses of other classes.
Information about the Wikidata ontology is not separated from the rest of the information in Wikidata, so Douglas Adams being an instance of the class human is represented by the statement
\hitem{https://www.wikidata.org/wiki/Q5}{Douglas Adams}{Q42} 
\hitem{https://www.wikidata.org/wiki/Property:P31}{instance of}{P31}
\hitem{https://www.wikidata.org/wiki/Q5}{human}{Q5}
and human being a subclass of person is 
\hitem{https://www.wikidata.org/wiki/Q5}{human}{Q5} 
\hitem{https://www.wikidata.org/wiki/Property:P279}{subclass of}{P279}
\hitem{https://www.wikidata.org/wiki/Q215627}{person}{Q215627}.
The Wikidata ontology includes classes at very different granularities,
ranging from specific classes such as 
\hitem{https://www.wikidata.org/wiki/Q271148}{monarch of Italy}{Q271148}
through classes like
\hitem{https://www.wikidata.org/wiki/Q5}{human}{Q5} 
to very general classes such as 
\hitem{https://www.wikidata.org/wiki/Q4406616}{concrete object}{Q4406616}
and even 
\hitem{https://www.wikidata.org/wiki/Q35120}{entity}{Q35120}, the universal class.

The Wikidata ontology treats classes just like any other item in Wikidata so classes can be instances of other classes.  For example, 
\hitem{https://www.wikidata.org/wiki/Q271148}{monarch of Italy}{Q271148}
is an instance of
\hitem{https://www.wikidata.org/wiki/Q5737899}{hereditary title}{Q5737899}
and 
\hitem{https://www.wikidata.org/wiki/Q4406616}{concrete object}{Q4406616}
 is an instance of 
\hitem{https://www.wikidata.org/wiki/Q104086571}{first-order class}{Q104086571}.

Wikidata does not have a formal basis, so there is no formal definition of 
\hitem{https://www.wikidata.org/wiki/Property:P31}{instance of}{P31}
and
\hitem{https://www.wikidata.org/wiki/Property:P279}{subclass of}{P279}.
Instead, there is general agreement of the intended meaning of items and properties in Wikidata, often embodied in item descriptions in Wikidata, such as ``this item is a subclass (subset) of that item'' for \hitem{https://www.wikidata.org/wiki/Property:P279}{subclass of}{P279}.

Wikidata also does not compute the consequences of these intended meanings, even for statements related to its ontology,
although for some intended meanings, such as inverses of properties, there are facilities in Wikidata that produce warnings when a consequence is missing from Wikidata. 
Instead users of Wikidata have to determine these consequences on their own.
For example, the \hitem{https://www.wikidata.org/wiki/Property:P279}{subclass of}{P279} property is transitive and an
\hitem{https://www.wikidata.org/wiki/Property:P31}{instance of}{P31}
a subclass of a class is also an instance of the class.
So users who want to determine the instances of a class using SPARQL have to write the query to also take into account subclasses, as in the following query that retrieves the instances of \hitem{https://www.wikidata.org/wiki/Q4406616}{concrete object}{Q4406616} and their English
labels:\footnote{Prefix declarations have been removed from queries as they are all standard for Wikidata queries.}
\begin{verbatim}
SELECT ?o ?oLabel WHERE {
  ?o wdt:P31/wdt:P279* wd:Q4406616.
  OPTIONAL { ?o rdfs:label ?oLabel.
     FILTER ( lang(?oLabel) = 'en' ) } }
\end{verbatim}

Wikidata is, out of necessity, not designed to have complete information and
missing information can be added at any time.
When examining Wikidata, this incompleteness has to be considered at all times.

Wikidata can be edited by anyone, and has a large community of editors including several automated systems (called {\em bots}).  As a result, there are many errors in Wikidata.  Some of these errors are related to differing interpretations of the information in Wikidata, some are
due to incorrect assumptions made by editors, and some are just due to editors misunderstanding the meaning of properties in Wikidata, including \witem{instance of}{P31} and \witem{subclass of}{P279}, the fundamental properties of the Wikidata ontology.


The basic interface to Wikidata at \url{wikidata.org} does not provide querying facilities beyond text searching.
It is, however, possible to query Wikidata with SPARQL using an encoding of Wikidata in RDF.
There are several services that provide this ability:
\begin{itemize}
\item \url{https://query.wikidata.org} (the Wikidata Query Service) using Blazegraph \cite{BlazeGraph} against a feed from Wikidata with a delay usually under 1 minute,
\item \url{https://wikidata.demo.openlinksw.com} using Virtuoso \cite{Virtuoso} against an RDF dump of Wikidata, updated irregularly, and
\item {https://qlever.cs.uni-freiburg.de/wikidata} using QLever \cite{QLever} against the current RDF dump of Wikidata, updated weekly.
\end{itemize}

All three of these SPARQL engines have problems.
Blazegraph is by far the slowest, timing out on many non-simple queries.
Virtuoso is faster, but can run into errors in its internal processing of queries.
QLever is very often by far the fastest but is not quite a complete SPARQL 1.1 implementation and sometimes tries to allocate too much memory.

The QLever service is used throughout this paper because it is the fastest engine for complex queries.
Using QLever has permitted deeper analysis of the problems of the Wikidata ontology than before.
Careful crafting of queries was sometimes required to reduce memory consumption in QLever.
Most results use the Wikidata RDF dump as of 18 June 2024.
For more information on the data and queries used see
the {\em Technical Appendix}.

\section{Related Work}

This work came out of a long-term desire to use Wikidata as a general-purpose source of information, requiring the ability to navigate the Wikidata ontology to find relevant information.  Because of the problems in the ontology actual efforts had to be limited to particular domains in Wikidata where the ontology was determined to be of sufficient quality or the problems could be overcome.

The problems in the Wikidata ontology are well known \cite{wikidata-ontology-issues-prioritization}, including not just problems related to class order but general problems with incorrect subclass and instance links in the ontology.
Several analyses of problems \cite{wikidata-ontology-design,wikidata-barriers,wikidata-semantic-web}
mostly present examples of different types of problems in the Wikidata ontology,
such as non-uniform modelling,
incorrect instance and subclass links,
vague descriptions of classes, and
semantic drift along subclass chains,
and make suggestions on how they might be fixed.
Other work \cite{wikidata-quality-study}
examines changes related to \witem{instance of}{P31} and \witem{subclass of}{P279} links,
concluding from the large number of changes of this type that the two properties are often misused.

Wikimedia Deutschland sponsored a series of events related to the ontology issues in Wikidata leading to a survey asking users what they though of various issues.   The results of this survey were published
\cite{wikidata-ontology-survey-prioritization, wikidata-ontology-issues}
and some potential solutions proposed.
\cite{wikidata-ontology-issues-solutions}
Even with these events and analyses not much actual effort has been made to fix the problems or to develop better technology for helping prevent problems.

Work by Brasileiro {\em et al.} (\citeyear{wikidata-multi-level-hierarchies})
has directly addressed the problems of multiple class orders
as needed in multi-level modeling,
observing that dealing with multiple levels of classification is conceptually difficult
and that there are modeling problems in the Wikidata ontology due to the improper use of the instance and subclass of properties violating their theory of class order.
The work presents and counts the occurrences of the anti-patterns of
a class being both an instance and a
subclass of another class
(15,177 occurrences),
a class being a subclass of two other classes one of which is an instance of the other
(441 occurrences)
and 
a class being both an instance of an instance of another class and an instance of the other class (7,082 occurrences)
as of early 2016.
The work claims that this is evidence that there are widespread problems with class order in Wikidata.
This work depends, however, on the ability to provide a fixed order for all classes in Wikidata.

Follow-on work by Dadalto {\em et al.} (\citeyear{wikidata-conceptual-disarray}) re-determined the prevalence of these anti-patterns
finding that over five years later their prevalence has increased dramatically,
to 2,035,434 occurrences of the first anti-pattern over a variety of domains but concentrated in biological domains,
to over 3,022,698 occurrences of the second anti-pattern again over a variety of domains but concentrated in biological and professional domains.
Due to computational difficulties this work used a filtered dump of Wikidata restricted to items that have \witem{subclass of}{P279} links.
The analysis also counted anti-patterns related to Wikidata classes that were explicitly marked as being first-order classes, finding that a large fraction of the occurrences involved these classes, even though modeling in first-order classes should be easy.

The work in this paper is most closely related to the two previous works.
This work updates, generalizes, and extends those investigations into problems related to class order in Wikidata.

\section{Classes and Class Order}

One can think of a class in Wikidata as an item that has an instance, as an item that has a subclass or a superclass, or as an instance of \witem{class}{Q16889133}, the class of all classes.  As Wikidata is incomplete, these three conditions do not necessarily correspond in Wikidata.  In fact, none of them dominate any of the others.

\begin{tabular}{cccc}
  \hline
  \multicolumn{4}{c}{Class Counts} \\
  \multicolumn{1}{c}{Has an} &
  \multicolumn{1}{c}{Has a sub-} &
  \multicolumn{1}{c}{Instance of} &
  \multicolumn{1}{c}{Any of} \\
  \multicolumn{1}{c}{instance} &
  \multicolumn{1}{c}{or superclass} &
  \multicolumn{1}{c}{\witems{class}{Q16889133}} &
  \multicolumn{1}{c}{these} \\
  \hline
  115,360  
  & 4,175,095 
  & 17,652,566 
  & 19,299,681 \\  
  \hline
\end{tabular}

\noindent
The number of instances of \witem{class}{Q16889133} seems suspiciously high.
This is investigated further later in the paper.

Class order is a useful way of naturally categorizing classes and is used in many places, including 
\cite{wikidata-multi-level-hierarchies, wikidata-conceptual-disarray}.
The definition of class order depends on what is a class and in several places in this paper several of the above conditions will be used.

A {\em first-order class} has no classes as instances.
A {\em second-order class} has only first-order classes as instances.
An {\em nth-order class} has only $n$-1th-order classes as instances.
Any non-empty class can have at most one (fixed) order.
A {\em variable-order class} has as instances classes of different orders or both non-classes and classes
Most classes should have a fixed order, but a few, like \witem{entity}{Q3150} the universal class, will be variable-order classes.

Here are some classes that should have fixed order:
\begin{tabular}{@{}l@{\hspace{0.25em}}l@{}}
\hline
First order&\witem{human}{Q5}, \witem{ship}{Q11446}, \\
	& \witem{iPad Air (5th generation)}{Q111154825} \\
Second order&\witem{watercraft type}{Q16335899}, \\
	& \witem{watercraft class}{Q18758641}, \witem{iPad}{Q2796}\footnotemark \\
Third order&\witem{model series}{Q811701} \\
\hline
\end{tabular}
\footnotetext{It probably would be better to label this as ``iPad model''.}

\section{Direct Determination of Class Order}

Wikidata has several built-in classes whose intended meaning is the class of all classes with a certain fixed order.\footnote{Wikidata has other classes related to class order, such as 
\witem{variable-order class}{Q23958852},
but these classes are not important for determining the precise fixed order of a class.}
These are
\witem{first-order class}{Q104086571},
\witem{second-order class}{Q24017414},
\witem{third-order class}{Q24017465},
\witem{fourth-order class}{Q24027474}, and
\witem{fifth-order class}{Q24027515}.
Each of these classes have fixed class order one higher than the order of their instances, i.e., \witem{second-order class}{Q24017414} is a third-order class, and has an \witem{instance of}{P31} link to the next-higher-order universal class.

From these five universal fixed-order classes it is possible to determine Wikidata classes that are required to have a fixed order using the following rules:
\begin{itemize}
\item If a class is an instance of a class of fixed order $n+1$ then it has fixed order $n$.
\item If a class is a subclass of a class of fixed order $n$ then it has fixed order $n$.
\end{itemize}
It is thus possible to determine the first order of many classes in Wikidata by appropriate queries using one or more of these classes.

It turns out that there were no subclasses of \witem{fifth-order class}{Q24027515}
and its only instance was \witem{fourth-order class}{Q24027474}.
Further there were no subclasses of \witem{fourth-order class}{Q24027474}
and its only instance was \witem{third-order class}{Q24017465}. 
Finally there were no subclasses of \witem{third-order class}{Q24017465}. 
As a result there were no classes in Wikidata that had a fixed order greater than three specified by their relationship to the universal fixed-order classes aside from universal fixed-order classes.  This simplifies the queries for the lower orders.

The third-order classes can be determined by the query:\footnote{The queries actually used also returned English labels.%
}
\begin{verbatim}
SELECT DISTINCT ?third WHERE {
  ?third wdt:P279*/wdt:P31 wd:Q24017465 . }
\end{verbatim}

\noindent
The second-order ones can be determined a similar query.
The first-order ones can be determined by the following query, which has been modified to reduce its memory consumption in QLever:
\begin{verbatim}
SELECT DISTINCT ?first WHERE {
  { SELECT DISTINCT ?c WHERE {
    ?c wdt:P31/wdt:P279*/wdt:P31/
        wdt:P279*/wdt:P31 wd:Q24017465. } }
  ?first wdt:P279* ?c. }
\end{verbatim}

There were 3,238 third-order,  
2,447,483 second-order, 
and
14,062,244 first-order classes 
returned by these queries.


These numbers seem high, based on the number of classes in Wikidata determined above and the number of classes that appear to contain non-classes.
But it might just be that there are many higher-order classes in Wikidata.

Looking at commonalities between these results, there are
3,159 third-order classes, out of 3,238, that are also second-order classes;
3,097 third-order classes, out of 3,238, that are also first-order classes; and
2,386,595 second-order classes, out of 2,435,283, that are also first-order classes.
As no non-empty class can have multiple fixed orders,
these numbers indicate that there are major errors in this part of Wikidata.
These errors are further examined later in this paper.

\section{Bounding of Class Order}

The above investigation ignores information about the instances of a class in Wikidata.  This information can be used to bound the order of a class, if it has one.

An {\em instance chain} is a sequence of items where each item is an instance of the previous one.  
If there is an instance chain of length $n$ in Wikidata starting at an item, then the class order of that item, if it has one, is at least $n$, independent of any other information about the item.
(Because of the interaction between instance of and subclass of
determining instance chains in Wikidata is not just looking for chains of \witem{instance of}{P31} links.)

It is possible to use instance chains to produce tighter bounds on class order by examining the item at the end of the chain.  For example, if there is an instance chain of length $n$ in Wikidata starting at an item and ending at a class, then the class order of that item, if it has one, is at least $n+1$.   (If the class at the end of the chain is a second-order class then the order is at least $n+2$, and so on, but this additional tightening has not been pursued.)

SPARQL is not suitable for directly determining the longest instance chains in Wikidata for each class.  It is, however, possible to create SPARQL queries that return classes with instance chains of length $n$ that start at the class and end at a class and thus the class has fixed order at least $n+1$ if it has a fixed order at all.  In this case the class is said to have a (non-strict) {\em minimum order} of $n+1$.

These SPARQL queries are complex and resource intensive.  In some cases to get them to run in the QLever service there had to be subqueries so partial results were de-duplicated before joining with the next part of the query.  For example, here is the query that computes the results for order at least three, i. e., classes that have an instance that has an instance that is a class:

\begin{verbatim}
SELECT DISTINCT ?third WHERE { 
  { SELECT DISTINCT ?second WHERE { 
    { SELECT DISTINCT ?c WHERE {
      { SELECT DISTINCT ?first WHERE {
        { ?i wdt:P31 ?first.  } UNION
        { ?sub wdt:P279 ?first. } UNION
        { ?first wdt:P279 ?super. } UNION
        { ?first wdt:P31/wdt:P279*
               wd:Q16889133. } } }
      ?first wdt:P31 ?c . } }
    ?c wdt:P279* ?second. } }
  ?second wdt:P31/wdt:P279* ?third. }
\end{verbatim}

The results for each minimum order computed are given in the following table:

\begin{tabular}{crrr}
\hline
Minimum & \multicolumn{1}{c}{Full} & \multicolumn{1}{c}{Query} & \multicolumn{1}{c}{Instance-Only}  \\
Order & \multicolumn{1}{c}{Count} & \multicolumn{1}{c}{Time} & \multicolumn{1}{c}{Count}  \\
\hline
1 & 19,299,681 &  7.0s & 115,360 \\  
2 &     44,716 & 12.9s &  13,847 \\  
3 &      7,678 & 13.6s &   3,647 \\  
4 & 	 2,854 & 13.6s &   1,924 \\  
5 &	 1,657 & 14.1s &   1,444 \\  
6 &	 1,413 & 14.4s &   1,345 \\  
\hline
\end{tabular}

\noindent
The second column is the count when using the full definition of a class.  The fourth column is the count when using the definition of a class as an item that has an instance.  The third column is the query time for the full count using QLever.


Many of these numbers are suspiciously high. As there are no fourth- or higher-order classes determinable by relationships to the universal fixed-order classes, it is unlikely that there are any classes in Wikidata whose fixed order should be four or more.
In particular, some classes that should be first order, such as 
\hitem{https://www.wikidata.org/wiki/Q5}{human}{Q5}, belong to each of sets above because they have an instance that is an instance of itself.

\section{Classes That Cannot Have Fixed Order}

A {\em split-order class} ({\sl S})
is a class where there is an item {\sl A} such that
{\sl A} is both a subclass of {\sl S} and a subclass of an instance of {\sl S}.
Both of the subclass relationships can be indirect or trivial.
A split-order class ({\sl S}) cannot have a fixed order because if {\sl S} had a fixed order the order would have to be both the same as and one more than the order of {\sl A}, if {\sl A} even had a fixed order.
For example, \witem{personalization}{Q1000371} is a subclass of \witem{knowledge system}{Q105948247}  and also an instance of \witem{specialty}{Q1047113} which is a subclass of \witem{knowledge system}{Q105948247}, so \witem{knowledge system}{Q105948247}
is a split-order class.
This criterion generalizes the anti-patterns of \cite{wikidata-conceptual-disarray}.

The query that finds all these pairs is 
\begin{verbatim}
SELECT DISTINCT ?c ?s WHERE {
  ?c wdt:P279* ?s . 
  ?c wdt:P279*/wdt:P31/wdt:P279* ?s . }
\end{verbatim}
\noindent
but this query has a possibly-empty path in it, which is not supported in QLever.
Several queries were constructed that covered the results of this query and ran successfully in the QLever service.
One of these queries is
\begin{verbatim}
SELECT DISTINCT ?c ?s WHERE {
  ?c wdt:P279+ ?s . 
  ?c wdt:P31/wdt:P279+ ?s . }
\end{verbatim}
\noindent
which returned 75,067,543 pairs. 
Combining the results of all the queries resulted in 84,158,269 pairs.  

It is useful to exclude items
that have where the item has a superclass with that is also related to the other class
or where the other class has a subclass that the item is related to, 
as fixing the issue for the reduced pair will fix the issue for original pair.
Queries were constructed to return these subsidiary classes and their results removed.
That still left 1,919,685 results for the query above  
and 3,821,300 results for all the cases, a substantial number.  


The results involved only 6,379 distinct split-order classes, however,
giving evidence that nearly all classes in Wikidata should be fixed order classes.
Some of the split-order classes show up in many pairs, indicating that they are not fixed-order classes.
For example, \witem{entity}{Q35120} shows up the most times, as expected.
But some classes that show up many times are not expected, including 
\witem{gene}{Q7187} and \witem{Q12136}{disease}.
Several prominent classes that should have a fixed order also show up,
including \witem{human}{Q5} and \witem{ship}{Q11446}.

\section{Other Ways to Determine Class Order}

There are several other, minor ways to determine class order.
The two described here could have been used to find more potential problems
but the above results indicate the presence of many problems
so that it didn't seem to be helpful to include these minor additions.

The description for \witem{metasubclass of}{P2445} says that it is a ``relation between two metaclasses: instances of this metaclass are likely to be subclasses of classes that are instances of the target metaclass''.   The slightly unusual wording of this description appears to be to allow for the possibility that no appropriate instance of the target metaclass has been added to Wikidata, even though it does exist.
From this description the subject metaclass should have the same class order of the object (target) metaclass, if it has one.

There were only 64 statements for this property.
Upon investigating them all at most 10 are valid, with most of the problems being related to class order. 
After publicizing the problematic statements to the Wikidata community they were removed using QuickStatements \cite{quickstatements}, a change tool for Wikidata,
on 23 July 2024.

The description for \witem{is metaclass for}{P8225} says that ``all instances of the subject are subclasses of the object''.  As a result, the subject should have class order one greater than the order of the object metaclass, if it has one.  

There were 787 statements for this property.
Most of them appear to be correct on a quick inspection 
but class orders cannot be deduced from them,
given the potential class-order problems already uncovered.


\section{Analysis and First Fixes}

Many of the problematic results above are too large to be investigated and fixed by hand by a small team of people.
Worse, the probable unreliability of all of the results means that one result cannot be taken as ground truth to help fix other results.   For example, a first-order class cannot have classes as instances but both of these results appear problematic so one cannot be used to help improve the other.
There will have to be deeper investigation to find root causes of the problems or find parts of the large results that can be trusted.

But there are a few results, or parts of results, that are small enough that manual fixing can be done in a reasonable amount of time and the effects of these changes determined to see whether small changes can produce dramatic improvements.  Several of these small results were examined and changes were made to fix or partially fix them.
Because edits were made to Wikidata itself\footnote{The major goal of this entire effort was, after all, to improve Wikidata.} rerunning queries also incorporated all the edits made to Wikidata over the intervening time and thus their results cannot be directly compared to the previous results.  Large effects, however, are very likely to be the result of the edits performed in this work.

\subsection{Direct Class-Order Conflicts}

There were 3,238 third-order classes with 3,158 of them also being second-order classes.  
Although this is not a trivial group it is small enough that it could be tackled by hand but it is more efficient to consider these classes starting with the most general.
There are only 51 roots 
(instances of \witem{third-order class}{Q24017465}) of the third-order classes
but 19 of them 
are also second-order classes. 

Many of these were in the chemistry domain, with a few in the manufactured objects domain.
Each were investigated, along with the other roots that might not be third-order, and the following changes 
made on 23 July 2024.

\noindent
\begin{tabular}{@{}l@{\hspace{.25em}}l@{}}
  \hline
  \multicolumn{1}{c}{Class} & \multicolumn{1}{c}{Change} \\
  \hline
  \witem{group or class of strains}{Q75913269} & remove third-order \\
  \witem{film rendition of song}{Q124354051} & inherit second-order \\
  \witem{prosodic group}{Q120733671} & remove third-order \\
  \witem{torpedo model}{Q104758804} & inherit second-order \\
  \witem{rail motor coach class}{Q63040753} & inherit second-order \\ 
  \witem{aircraft family}{Q15056993} & change superclass \\ 
  \witem{combat aircraft family}{Q124054738} & change superclass \\ 
  \witem{combat vehicle family}{Q100709275} & change superclass \\ 
  \wname{group or class of chemical entities} & remove second- \\
  \witem{}{Q72070508} & order parent \\
  \hline
\end{tabular}

\noindent
The change with the most potential effect was removing the second-order parent of \witem{group or class of chemical entities}{Q72070508}.
This removed a source of conflict between third-order and second-order classes for each of the over 3,000 subclasses of this class.
Over three million other descendants of this class where also potentially affected.

All the third-order classes that were not in the chemistry domain were also investigated.
Only 19 of them were also second-class. 
Of these, several were incorrectly linked to \witem{second-order class}{Q24017414},
several were not third-order but were incorrectly linked to \witem{model series}{Q811701} or some other third-order class.  These errors were fixed using QuickStatements on 20 July 2024.

\subsection{Instance Loops}

An instance loop is a item that is an instance of itself and is a special case of split-order classes.
There were 120 items in instance loops, each of them was analyzed and 113 were found to be incorrect, 
for example \witem{Fabio Valentini}{Q94440081},
with only seven very general classes, such as \witem{entity}{Q35120}, correctly being instances of themselves.  
These 113 items, however, make 1,171 classes 
not be of fixed order because the class has an infinite descending chain of instances.  
These classes all also show up in each of the minimum-order levels.
Breaking these loops could reduce the number of classes with minimum order 4 or more dramatically and have a large effect on class order errors.

All 113 incorrect instance loops were broken by removing an incorrect instance or subclass link
on 19 July 2024.
For the 86 loops that involved a single link, that link was removed.
The other loops were examined and either an instance or subclass link in the loop was removed.
The changes were put into a single QuickStatements batch and performed on 19 July 2024.

A first analysis showed reduction of around one-half in the number of high-minimum-order classes but many remained.
The problem is that there where instance loops that were not detected in the initial queries.

Some of these were caused by instance loops with two instance links,
for example where an item is an instance of another item which is in turn an instance of the first,
as in \witem{legal abuse}{Q6517445}
was an instance of \witem{abuse of rights}{Q12586068}
which was, in turn, an instance of \witem{legal abuse}{Q6517445}.  
All of these loops were investigated.
Some of these were correct but many were incorrect and were fixed on 29 July 2024.  

Some of the remaining too-high minimum-order classes were also caused by incorrect statements that did not involve loops.
For example, there were 591 non-looping direct instances of instances of \witem{human}{Q5}
such as \witem{Bennifer}{Q108419447} being an instance of \witem{Jennifer Lopez}{Q40715}.  
Of these, 538 were caused by \witem{Ogham Person Concept}{Q110897921} incorrectly being an instance of \witem{human}{Q5} instead of a subconcept.
Of the remainder, most were items, like books, that were incorrectly instances of actual humans.
Fifty-five changes were made on 29 July 2024 to fix these problems. 

\subsection{Split-Order Classes}

There are 6,379 split-order classes, which cannot have a fixed order.  
Of these, 3,037 are split-order because of only one item, and 5,346 because of ten or fewer.
Most of these are probably an error and should be fixed but examining all these items would be a lengthy task.

Nevertheless, several single-item split-order classes were investigated and the items adjusted to make the class no longer split-order using a QuickStatements batch on 22 July 2024.
A few prominent split-order classes with few items---\witem{human}{QA5}, \witem{mammal}{Q110551885}, \witem{animal}{Q279}, \witem{mountain}{Q8502}, \witem{lake}{Q23397}---were also investigated and their items adjusted as well.
In most cases determination of what change could easily be made by looking at the item or looking at the statements involving the split-order class.
These spot fixes will not make major changes to the number of split-order classes
so the queries were not rerun.

There are some classes that have many items causing their split order.  Here are the classes with the highest numbers:

\begin{tabular}{rl}
\hline
\multicolumn{1}{c}{Count} & \multicolumn{1}{c}{Class} \\
\hline
1,497,615 & \witem{entity}{Q35120} \\
975,522 & \witem{gene}{Q7187} \\
761,343 & \witem{protein}{Q8054} \\
55,883 & \witem{artificial object}{Q16686448} \\
49,676 & \witem{product}{Q2424752} \\
49,392 & \witem{pseudogene}{Q277338} \\
45,361 & \witem{collective entity}{Q99527517} \\
45,201 & \witem{non-coding RNA}{Q427087} \\
27,047 & \witem{abstract entity}{Q7048977} \\
19,973 & \witem{object}{Q488383} \\
17,943 & \witem{continuant}{Q103940464} \\
16,125 & \witem{physical object}{Q223557} \\
15,720 & \witem{part}{Q15989253} \\
14,921 & \witem{class}{Q16889133} \\
11,806 & \witem{position}{Q4164871} \\
10,141 & \witem{goods}{Q28877} \\
9,162 & \witem{work}{Q386724} \\
9,087 & \witem{CN}{Q22231119} \\
6,384 & \witem{human language}{Q20162172} \\
6,005 & \witem{language}{Q34770}  \\
\hline
\end{tabular}

\noindent
Some of these are truly variable order, like \witem{entity}{Q35120} and \witem{class}{Q16889133}.
But some appear to be fixed-order classes, like \witem{gene}{Q7187} and \witem{product}{Q2424752}.
Each of these classes need to be investigated to determine their status, probably in discussion with the portion of the Wikidata community working in the area.  For those classes that are determined to have a fixed order
hopefully semi-automated methods specific to the classes can be developed to fix the splits causing order problems for the class.

\subsection{Large Number of Classes}

Although not exactly a class order problems, the very large number (17,652,566)  
of instances of \witem{class}{Q16889133} suggests the presence of incorrect information.

The major direct subclasses of \witem{class}{Q16889133} and their number of instances are:

{\small
\begin{tabular}{rl}
\hline
  \multicolumn{1}{c}{Count} & \multicolumn{1}{c}{Subclass} \\
\hline
6,110,298 & \witem{list}{Q12139612} \\
4,031,120 & \witem{group or class of organisms}{Q21871294} \\
3,323,351 & \witem{sequence}{Q20937557} \\
2,014,525 & \witem{type}{Q21146257} \\
782,246 & \witem{collection}{Q2668072} \\
735,211 & \witem{version,  edition or translation}{Q3331189} \\
126,082 & \witem{convention}{Q367293} \\
101,484 & \witem{technical standard}{Q317623} \\
63,064 & \witem{metaclass}{Q19478619} \\
48,167 & \witem{group or class of proteins}{Q84467700} \\
46,677 & \witem{legal form}{Q10541491} \\
44,612 & \witem{protein family}{Q417841} \\
\end{tabular}
\begin{tabular}{rl}
33,264 & \witem{vegetational formation}{Q2083910} \\
30,235 & \witem{genre}{Q483394} \\
28,468 & \witem{syntaxon}{Q2471463} \\
23,776 & \witem{brand}{Q431289} \\
19,105 & \witem{abstract class}{Q122375923} \\
17,664 & \witem{fixed-order class}{Q23959932} \\
17,080 & \witem{first-order class}{Q104086571} \\
12,733 & \witem{dyad}{Q29431432} \\
\hline
\end{tabular}
}

The very largest and some of the other very large groups of classes have to do with lists and other sequence-related concepts.\footnote{Note that the number of instances of \witem{fixed-order class}{Q23959932} and \witem{first-order class}{Q104086571} do not include subclasses of the instances and thus are severe undercounts.}
Although Wikidata considers these to be classes, there is a case to be made that they are not really classes.  Changing the situation would require a sustained discussion with the Wikidata community.

\section{Resulting Counts}

After the above changes all showed up in a Wikidata dump and were available in QLever,
the queries for order conflicts and minimum order were rerun.


The class order counts and conflicts changed as follows:

\noindent
{\small
\begin{tabular}{@{}c@{}r@{\hspace{0.25em}}r@{\hspace{0.25em}}r@{\hspace{0.25em}}r@{\hspace{0.25em}}r@{\hspace{0.25em}}r@{}}
\hline
Or- & \multicolumn{2}{c@{}}{Count} & \multicolumn{2}{c@{}}{Order 2 Conflicts}   & \multicolumn{2}{c@{}}{Order 1 Conflicts}  \\
der & \multicolumn{1}{c@{}}{Initial} & \multicolumn{1}{c@{}}{Revised} & \multicolumn{1}{c@{}}{Initial} & \multicolumn{1}{c@{}}{Revised} & \multicolumn{1}{c@{}}{Initial} & \multicolumn{1}{c@{}}{Revised}  \\
\hline
3 & 3238 & 3565 & 3159 & 3446 & 3097 & 3435 \\
2 & 2447483 & 2436826 &   &   & 2386595 & 2424609 \\
1 & 14062244 & 13185450 &   &   &   &  \\
\hline
\end{tabular}
}

\noindent
Given that a source of conflict was removed for over 3,000 third-order classes in the chemical domain by separating them from a class in the physical domain, these results are disappointing.
Looking closer into these classes uncovers several class-order conflicts within biochemistry.
In particular, \witem{protein family}{Q417841} is directly a second-order class but also
a third-order class via \witem{group or class of proteins}{Q84467700}.
Further investigation into this domain is needed to eliminate the conflicts either by manually going through all the classes or by writing new queries to find sources for conflicts.

The minimum-order counts changed as follows:
\noindent
\begin{tabular}{@{}c@{}r@{\hspace{0.25em}}r@{\hspace{0.25em}}r@{\hspace{0.25em}}r@{\hspace{0.25em}}r@{\hspace{0.25em}}r@{}}
\hline
Minimum & \multicolumn{2}{c}{Full} & \multicolumn{2}{c}{Instance-Only}  \\
Order & \multicolumn{1}{c}{Initial} & \multicolumn{1}{c}{Revised} & \multicolumn{1}{c}{Initial} & \multicolumn{1}{c}{Revised}  \\
\hline
1 & 19,299,681 & 19,025,409 & 115,360 & 115,467 \\
2 &     44,716 &  44662 & 13,847 & 13,661 \\
3 &      7,678 &   7,552 & 3,647 & 3,346 \\ 
4 & 	 2,854 &   1,924 & 2,430 & 1,411 \\ 
5 &	 1,657 &   1,065 & 1,444 & 794 \\
6 &	 1,413 &   722 & 1,345 & 591 \\
\hline
\end{tabular}

\noindent
The minimim-order counts reduced significantly for orders four and above, indicating that many class order issues were caused by instance loops.  But many high-mimimum-order classes remain, indicating that further exploration is needed.

\section{Suggestions}

One driver of class order problems is that
editors never have to confirm their edit if it causes a problem,
problems are often not shown at all to editors at edit time,
and for many problems there are not even reports generated to show the problem.
This is partly due to hostility to anything that makes it harder to edit Wikidata,
even if the edits violate community norms or intended meaning of Wikidata constructs,
and partly due to a lack of tools that can compute class order problems.
As a result new problems are continually introduced, often inadvertently.
But, for example, if users were shown a caution when they create an instance loop this simple source of problems could be considerably reduced.

One way of generating better reports would be to move the Wikidata Query Service from the very slow Blazegraph
to the much faster QLever.  This requires enhancements to QLever, as QLever has not had a way of removing triples from the graph it is querying.
Providing more information to editors requires changes to tools that are used to edit Wikidata,
so that class order and more other problems are found and shown prominently,
ideally requiring confirmation when an edit causes a problem.
A bigger change would be to have a larger portion of the Wikidata community
emphasizing improvements to Wikidata over additions,
making it more like English Wikipedia in this respect.

Another social change for Wikidata would be to put more effort into the Wikidata ontology community,
particularly to investigate how the Wikidata ontology is used in the other parts of the Wikidata community
and educate these sub-communities on how best to use the Wikidata ontology.
Some of the major improvements here were the result of fixing misalignments between how different communities use the Wikidata ontology, for example class order misalignments between chemistry and physical object
and between combat vehicles and vehicles in general.

Fixing existing class order problems might be best done by creating an interface
where users are shown problem instances and asked to determine a change to fix the problem.
The interface would have to be carefully constructed so that users understand what the problem is
and what kinds of fixes are appropriate.
Incentives could be set up to turn this activity into a kind of game,
as has been done in other crowd-sourced areas \cite{wikipedia-gamifying,gamifying-lexicographical}.

\section{Summary and Future Work}

This paper shows that it is possible using current tools (QLever) to
completely enumerate certain kinds of problems related to class order in Wikidata
A small number of edits to Wikidata can fix large fractions of some of these problems.
Other problems appear to require very many edits to fix.

The effort to improve the Wikidata ontology continues, with fixing class order problems a major part.
A next step is to continue the analysis of the results of the first set of fixes and
create queries that show more problems.
If the fixes make some of the results trustworthy they can be used to help fix other problems,
such as using the directly-determined first-order classes to drive changes to items that make them be split-order.
The queries used so far were almost all agnostic with respect to the meaning of classes (although the fixes did take class meanings into account).
Another next step is to take class meaning more into account, for example by noting that \witem{ concrete object}{Q4406616} is a first-order class so instances of instances of the class signal a problem that needs to be addressed.

\section{Acknowledgments}
The authors thank Hanna Bast and the rest of the QLever team for providing the QLever
Wikidata service and helping revise some queries to get around a QLever bug and 
reduce memory consumption.

\bibliography{order}

\appendix
\section{Technical Appendix}

The queries here were first run using the QLever Wikidata SPARQL service at \url{qlever.cs.uni-freiburg.de/wikidata}
around 20 June 2024
over an RDF graph containing the full Wikidata RDF dump files 
\url{20240617/wikidata-20240617-all-BETA.ttl.bz2}
and
\url{20240621/wikidata-20240621-lexemes-BETA.ttl.bz2}
from 
\url{dumps.wikimedia.org/wikidatawiki/entities/}.
Many of the queries were also run on 3 August 2024 
over an RDF graph containing the full Wikidata RDF dump files 
\url{20240729/wikidata-20240729-all-BETA.ttl.bz2}
and
\url{20240802/wikidata-20240802-lexemes-BETA.ttl.bz2}.
The dump files are only available for a few months after first availability.
Archived copies will be available by contacting the first author.
Some queries had to be run using a different QLever server that had a longer timeout for downloads.

A machine comparable to a Ryzen 9 7900X with 128GB of memory is adequate to run the queries.
The results of many of the queries are much too long to be included in an appendix, as some of them are over 2GB in size, but will be available by contacting the first author.

All queries use prefixes standard when querying Wikidata, namely

\begin{tabular}{ll}
wd &   http://www.wikidata.org/entity/ \\
wdt &  http://www.wikidata.org/prop/direct/ \\
rdfs & http://www.w3.org/2000/01/rdf-schema\#
\end{tabular} \\
All queries also return item labels, using variations of 
\begin{verbatim}
OPTIONAL { ?c rdfs:label ?c .
           FILTER ( lang(?c) = 'en' ) }
\end{verbatim}
These parts of the queries are not shown here.

\subsection{Classes and Class Order}

\noindent
{\em Class Counts for ``Has an instance'':}
\begin{verbatim}
SELECT DISTINCT ?first WHERE {
 {SELECT DISTINCT ?c WHERE {?i wdt:P31 ?c.}}
 ?c wdt:P279* ?first .
}
\end{verbatim}

\noindent
{\em Class Counts for ``Has a sub- or superclass'':}
\begin{verbatim}
SELECT DISTINCT ?first WHERE {
 { ?sub wdt:P279 ?first . } UNION 
 { ?first wdt:P279 ?super . }
} 
\end{verbatim}

\noindent
{\em Class Counts for ``Instance of \witems{class}{Q16889133}'':}
\begin{verbatim}
SELECT DISTINCT ?first WHERE {
 ?first wdt:P31/wdt:P279* wd:Q16889133 .
}
\end{verbatim}

\noindent
{\em Class Counts for ``Any of these'':}
\begin{verbatim}
SELECT DISTINCT ?first WHERE {
 { ?i wdt:P31 ?first.  } UNION 
 { ?sub wdt:P279 ?first . } UNION
 { ?first wdt:P279 ?super . } UNION 
 { ?first wdt:P31/wdt:P279* wd:Q16889133 . }
} 
\end{verbatim}

\subsection{Direct Determination of Class Order}

Because there are no fourth-order or fifth-order classes except for universal fixed-order classes
and because each universal fixed-order class is an instance of the next-higher universal fixed-order classes simpler queries can be used to return the third-order, second-order, and first-order classes.

\noindent
{\em Third-order classes:}
\begin{verbatim}
SELECT DISTINCT ?third WHERE {
  ?third wdt:P279*/wdt:P31 wd:Q24017465 .
}
\end{verbatim}

\noindent
{\em Second-order classes:}
\begin{verbatim}
SELECT DISTINCT ?second WHERE {
 ?second wdt:P279*/wdt:P31/wdt:P279*/wdt:P31
    wd:Q24017465. 
}
\end{verbatim}

\noindent
{\em First-order classes:}
\begin{verbatim}
SELECT DISTINCT ?first WHERE {
 { SELECT DISTINCT ?c WHERE {
  ?c wdt:P31/wdt:P279*/wdt:P31/
      wdt:P279*/wdt:P31 wd:Q24017465 .
 } }
 ?first wdt:P279* ?c .
}
\end{verbatim}

\noindent
The last query was modified to reduce memory consumption in QLever.
The commonalities between the results were computed using the Linux command {\tt comm} 
after removal of lines that do not contain results and then sorting.

\subsection{Bounding of Class Order}

Some of the queries here use subqueries to reduce memory consumption in QLever.

\noindent
{\em Order 1:}
See ``Queries for Classes'' above.

\noindent
{\em Order 6 instance-only:}
\begin{verbatim}
SELECT DISTINCT ?sixth WHERE { 
 { SELECT DISTINCT ?fifth WHERE { 
  {  SELECT DISTINCT ?fourth WHERE { 
   { SELECT DISTINCT ?third WHERE { 
    { SELECT DISTINCT ?second WHERE { 
     { SELECT DISTINCT ?first WHERE {
      { SELECT DISTINCT ?c WHERE {
         ?zero wdt:P31 ?c . } }
       ?c wdt:P279* ?first .
       } }
      ?first wdt:P31/wdt:P279* ?second .
      } }
     ?second wdt:P31/wdt:P279* ?third .
     } }
    ?third wdt:P31/wdt:P279* ?fourth .
    } }
  ?fourth wdt:P31/wdt:P279* ?fifth .
  } }
 ?fifth wdt:P31/wdt:P279* ?sixth .
}
\end{verbatim}

\noindent
{\em Order 6 full:}
\begin{verbatim}
SELECT DISTINCT ?sixth WHERE { 
 { SELECT DISTINCT ?fifth WHERE { 
  { SELECT DISTINCT ?fourth WHERE { 
   { SELECT DISTINCT ?third WHERE { 
    { SELECT DISTINCT ?second WHERE { 
     { SELECT DISTINCT ?c WHERE {
      { SELECT DISTINCT ?fr WHERE {
       {?i wdt:P31 ?fr.} UNION
       {?sub wdt:P279 ?fr.} UNION
       {?fr wdt:P279 ?super.} UNION
       {?fr wdt:P31/wdt:P279* wd:Q16889133.}
      } }
      ?fr wdt:P31 ?c .
     } }
     ?c wdt:P279* ?second .
    } }
    ?second wdt:P31/wdt:P279* ?third .
   } }
   ?third wdt:P31/wdt:P279* ?fourth .
  } }
  ?fourth wdt:P31/wdt:P279* ?fifth .
  } }
 ?fifth wdt:P31/wdt:P279* ?sixth .
}
\end{verbatim}

\noindent
{\em Order 2--5:}  Remove the outer query or queries in the above, as appropriate.

\subsection{Classes That Cannot Have Fixed Order}

The query that would compute all pairs of split-order classes (\verb|$s|) and items that make them split-order (\verb|$c|) is
\begin{verbatim}
SELECT DISTINCT ?c ?s WHERE {
  ?c wdt:P279* ?s . 
  ?c wdt:P279*/wdt:P31/wdt:P279* ?s .
}
\end{verbatim}
\noindent
This query cannot be evaluated in QLever due to a possibly-empty path
so it was split into caes and queries were constructed for each case.
The results of these queries were combined using the Linux command {\tt comm}, after removal of the first lines and then sorting, to combine the results.
For each of these queries, exclusion queries were constructed and subtracted using {\tt comm}.

In the cases below $\rightarrow$ is \witem{instance of}{P31} and
$\Rightarrow$ is \witem{subclass of}{P279}.

\noindent
{\em CASE self: C is the same as S, so C wdt:P31/wdt:P279* C}

\noindent
This query uses a {\tt FILTER} to get around a bug in QLever.
\begin{verbatim}
SELECT DISTINCT ?c WHERE {
 { ?c wdt:P31 ?c . } UNION
 { ?c wdt:P31/wdt:P279+ ?s .
   FILTER ( ?c = ?s ) }
}
\end{verbatim}
The query separates out two subcases of this case to get around a bug in QLever.

\noindent
{\em CASE other:  C is a proper subclass of S}

\noindent
The straightforward query runs out of memory in QLever so it was split into subcases using the fact that transitive-reflexive closure is the union of transitive closure and identity.

\noindent
{\em Subcase AB:  C wdt:P279*/wdt:P31 S}

\begin{verbatim}
SELECT DISTINCT ?c ?s {
 ?c wdt:P279+ ?s . 
 ?c wdt:P279*/wdt:P31 ?s .
}
\end{verbatim}

\noindent
{\em Subcase C:  C wdt:P31/wdt:P279+ S}
\begin{verbatim}
SELECT DISTINCT ?c ?s WHERE {
 ?c wdt:P279+ ?s . 
 ?c wdt:P31/wdt:P279+ ?s .
}
\end{verbatim}

\noindent
{\em Subcase D:  C wdt:P279+/wdt:P31/wdt:P279+ S}
\begin{verbatim}
SELECT DISTINCT ?c ?s WHERE {
  { SELECT DISTINCT ?c ?s WHERE {
     ?c wdt:P279+ ?s . } }
  { SELECT DISTINCT ?c ?s WHERE {
     ?c wdt:P279+/wdt:P31/wdt:P279+ ?s . } }
}
\end{verbatim}

\noindent
This query was modified to reduce memory consumption in QLever.

\noindent
{\em Exclusions:}

No exclusions are needed for the self case.  For each subcase of the other case, one or more queries that executed sucessfully in QLever were constructed for parts of the exclusion.  Some of these queries use subqueries to reduce memory consumption in QLever.

\noindent
{\em Subcase AB exclusion:}
\begin{verbatim}
SELECT DISTINCT ?c ?s WHERE {
 ?c wdt:P279 ?cx .
 ?cx wdt:P279+ ?s . 
 ?cx wdt:P279*/wdt:P31 ?s .
}
\end{verbatim}

\noindent
{\em Subcase C exclusion 1:}
\begin{verbatim}
SELECT DISTINCT ?c ?s WHERE {
 { SELECT DISTINCT ?c ?s WHERE {
  { SELECT DISTINCT ?c ?sx WHERE {
   ?c wdt:P279+ ?sx . 
   ?c wdt:P31/wdt:P279+ ?sx . } }
  ?sx wdt:P279 ?s .
 } }
}
\end{verbatim}

\noindent
{\em Subcase C exclusion 2:}
\begin{verbatim}
SELECT DISTINCT ?c ?s WHERE {
 { SELECT DISTINCT ?c ?s WHERE {
  { SELECT DISTINCT ?c ?sx WHERE {
   ?c wdt:P279+ ?sx . 
   ?c wdt:P31 ?sx . } }
  ?sx wdt:P279 ?s .
  } }
}
\end{verbatim}

\noindent
{\em Case D exclusion 1:}
\begin{verbatim}
SELECT DISTINCT ?c ?s WHERE {
 { SELECT DISTINCT ?c ?sx WHERE {
  { SELECT DISTINCT ?c ?sx WHERE {
     ?c wdt:P279+ ?sx . } }
  { SELECT DISTINCT ?c ?sx WHERE {
     ?c wdt:P279+/wdt:P31/wdt:P279+ ?sx. } }
 } }
 ?sx wdt:P279 ?s .
}
\end{verbatim}

\noindent
{\em Case D exclusion 2:}
\begin{verbatim}
SELECT DISTINCT ?c ?s WHERE {
 { SELECT DISTINCT ?c ?sx WHERE {
  { SELECT DISTINCT ?c ?sx WHERE {
      ?c wdt:P279+ ?sx . } }
  { SELECT DISTINCT ?c ?sx WHERE {
      ?c wdt:P279+/wdt:P31 ?sx . } }
 } }
 ?sx wdt:P279 ?s .
}
\end{verbatim}

\noindent
{\em Case D exclusion 3:}
\begin{verbatim}
SELECT DISTINCT ?c ?s WHERE {
 { SELECT DISTINCT ?cx ?s WHERE {
  { SELECT DISTINCT ?cx ?s WHERE {
     ?cx wdt:P279+ ?s . } }
  { SELECT DISTINCT ?cx ?s WHERE {
     ?cx wdt:P279+/wdt:P31/wdt:P279+ ?s.} }
 } }
 ?c wdt:P279 ?cx .
}
\end{verbatim}

\noindent
{\em Case D exclusion 4:}
\begin{verbatim}
SELECT DISTINCT ?c ?s WHERE {
 { SELECT DISTINCT ?cx ?s WHERE {
  ?cx wdt:P279+ ?s .
  ?cx wdt:P31/wdt:P279+ ?s .
 } }
 ?c wdt:P279 ?cx .
}
\end{verbatim}

\subsection{Other Ways to Determine Class Order}

\begin{figure*}
\begin{tabular}{ll}
\witem{product category}{Q63981612} & \witem{product}{Q2424752} \\
\witem{product model}{Q10929058} & \witem{model series}{Q811701} \\
\witem{aircraft model}{Q15056995} & \witem{aircraft family}{Q15056993} \\
\witem{weapon model}{Q15142894} & \witem{weapon family}{Q15142889} \\
\witem{combat aircraft model}{Q124054999} & \witem{combat aircraft family}{Q124054738} \\
\witem{artillery model Q124056256}{Q124056273} & \witem{family}{artillery} \\
\witem{firearm model}{Q22704163} & \witem{firearm family}{Q124054676} \\
\witem{ammunition model}{Q42314054} & \witem{ammunition family}{Q42314051} \\
\witem{torpedo model}{Q104758804} & \witem{torpedo family}{Q124054628} \\
\witem{missile model}{Q18487055} & \witem{missile family}{Q18487018} \\
\witem{combat vehicle model}{Q100710213} & \witem{combat vehicle family}{Q100709275} \\
\witem{bus model}{Q23039057} & \witem{bus family}{Q56697263} \\
\witem{vehicle family}{Q22999537} & \witem{automobile model}{Q3231690} \\
\witem{torpedo model}{Q104758804} & \witem{weapon model}{Q15142894} \\
\witem{torpedo model}{Q104758804} & \witem{weapon model}{Q15142894} \\
\witem{artillery model series}{Q124056266} & \witem{artillery family}{Q124056256} \\
\witem{ammunition model series Q42314051}{Q124054879} & \witem{family}{ammunition} \\
\witem{firearm model series}{Q124054897} & \witem{firearm family}{Q124054676} \\
\witem{missile model series}{Q124054933} & \witem{missile family}{Q18487018} \\
\witem{torpedo model series}{Q124054944} & \witem{torpedo family}{Q124054628} \\
\witem{combat vehicle model series}{Q124054888} & \witem{firearm family}{Q124054676} \\
\witem{combat aircraft model series}{Q124054994} & \witem{combat aircraft family}{Q124054738} \\
\witem{History by subject in Saudi Arabia}{Q105575131} & \witem{type}{Q21146257} \\
\witem{History by subject in Saudi Arabia}{Q105575128} & \witem{topic}{Q200801} \\
\witem{History by time in Saudi Arabia}{Q11471} & \witem{time}{Q105575122} \\
\witem{History by place in Saudi Arabia}{Q105575125} & \witem{physical location}{Q17334923} \\
\witem{History by event in Saudi Arabia}{Q105575134} & \witem{event}{Q10290214} \\
\witem{heritage of Saudi Arabia}{Q105582463} & \witem{heritage designation}{Q30634609} \\
\witem{heritage of Saudi Arabia}{Q105582463} & \witem{historic preservation}{Q914856} \\
\witem{budget label}{Q1001388} & \witem{type of record label}{Q107317889} \\
\witem{reissue label}{Q107317983} & \witem{type of record label}{Q107317889} \\
\witem{jazz label}{Q107318520} & \witem{type of record label}{Q107317889} \\
\witem{heavy metal label}{Q111450726} & \witem{type of record label}{Q107317889} \\
\witem{reggae label}{Q114707507} & \witem{type of record label}{Q107317889} \\
\witem{punk label}{Q115548417} & \witem{type of record label}{Q107317889} \\
\witem{Wikimedia list of musical works by performer}{Q107428796} & \witem{Wikimedia list of musical works}{Q100775261} \\
\witem{soundtrack album released on this record label}{Q61338340} & \witem{album released on this record label}{Q29033193} \\
\witem{soundtrack album by this artist Q29035150}{Q61507380} & \witem{by this artist}{album} \\
\witem{compilation album by this artist}{Q61511035} & \witem{album by this artist}{Q29035150} \\
\witem{remix album by this artist}{Q61565021} & \witem{album by this artist}{Q29035150} \\
\witem{musical work by this composer}{Q82775187} & \witem{work by this creator}{Q106796414} \\
\witem{musical work by this composer}{Q82775187} & \witem{work by this person}{Q111846649} \\
\witem{jazz musician playing this instrument}{Q66712100} & \witem{instrumentalist playing this instrument}{Q66712069} \\
\witem{works written by this person}{Q111590755} & \witem{item by creator}{Q106913988} \\
\witem{drug class}{Q2585617} & \witem{group or class of chemical substances}{Q17339814} \\
\witem{group of chemical entities}{Q55640599} & \witem{group or class of chemical entities}{Q72070508} \\
\witem{Category:Gords by country}{Q116767711} & \witem{gord}{Q88598} \\
\witem{Category:Gords in the Czech Republic}{Q8915887} & \witem{Category:Gords by country}{Q116767711} \\
\witem{dialect of a file format}{Q57696248} & \witem{file format family}{Q26085352} \\
\witem{Pal element}{Q124413096} & \witem{class of fictional entities}{Q15831596} \\
\witem{Pokémon type}{Q1266830} & \witem{class of fictional entities}{Q15831596} \\
\witem{battery format}{Q4374872} & \witem{type of object}{Q96251598} \\
\witem{extinct taxon}{Q98961713} & \witem{taxon}{Q16521} \\
\witem{mineral supergroup}{Q3977918} & somevalue
\end{tabular}
\caption{Removed \witem{metasubclass of}{P2445} statements}
\label{metasubclass-removed}
\end{figure*}

The \witem{metasubclass of}{P2445} statements in Figure 
\ref{metasubclass-removed} were removed.
The following \witem{metasubclass of}{P2445} statements were retained:

\noindent
\begin{tabular}{@{}l@{\hspace{0.5em}}l@{}}
\witem{mineral species}{Q12089225} & \witem{mineral class}{Q12110334} \\
\witem{mineral subclass}{Q3965272} & \witem{mineral class}{Q12110334} \\
\witem{mineral group}{Q1936581} & {\em mineral supergroup} \\
	 & \witem{}{Q3977918} \\
\witem{mineral variety}{Q12089225} & \witem{mineral species}{Q429795} \\
\witem{dog variety}{Q15852455} & \witem{dog breed}{Q39367} \\
\witem{order}{Q1865678} & \witem{class}{Q13744019} \\
\witem{alliance}{Q12057445} & \witem{order}{Q1865678} \\
\witem{suballiance}{Q4812777} & \witem{alliance}{Q12057445} \\
\witem{association}{Q744745} & \witem{alliance}{Q12057445} \\
\witem{subassociation}{Q4354416} & \witem{association}{Q744745}
\end{tabular}

\subsection{Analysis and First Fixes}

\subsubsection{Direct Class-Order Conflicts} ~ \\

\noindent
As the changes to third-order classes had the most impact both queries and results are given here.

\vspace{0.5ex}
\noindent
{\em Third-order roots:}
\begin{verbatim}
SELECT DISTINCT ?third WHERE {
  ?third wdt:P31 wd:Q24017465 .
}
\end{verbatim}

\noindent
{\em Third-order roots initial result, and determined class order:}

\noindent
{\small
\begin{tabular}{ll}
\witem{second-order class}{Q24017414} & 3rd \\
\witem{model series}{Q811701} & 3rd \\
\witem{firearm model series}{Q124054897} & 3rd \\
\witem{missile model series}{Q124054933} & 3rd \\
\witem{torpedo model series}{Q124054944} & 3rd \\
\witem{artillery model series}{Q124056266} & 3rd \\
\witem{combat aircraft model series}{Q124054994} & 3rd \\
\witem{weapon model series}{Q115044473} & 3rd \\
\witem{ammunition model series}{Q124054879} & 3rd \\
\witem{combat vehicle model series}{Q124054888} & 3rd \\
\witem{vehicle model series}{Q29048319} & 3rd \\
\witem{tractor model series}{Q55901595} & 3rd \\
\witem{automobile model series}{Q59773381} & 3rd \\
\witem{motorcycle model series}{Q71310524} & 3rd \\
\witem{scientific instrument model series}{Q117802245} & 3rd \\
\witem{rocket series}{Q111722634} & 3rd \\
\witem{engine series}{Q118017625} & 3rd \\
\witem{torpedo model}{Q104758804} & 2nd \\
\witem{rail motor coach class}{Q63040753} & 2nd \\
\witem{group or class of chemical entities}{Q72070508} & 2nd \\
\witem{class of chemical entities with \ldots}{Q56256173} & 2nd \\
\witem{class of chemical entities with \ldots}{Q56256178} & 2nd \\
\witem{imprecise class of chemical entities}{Q74892521} & 2nd \\
\witem{group of chemical entities}{Q55640599} & 2nd \\
\witem{group of isomeric entities}{Q15711994} & 2nd \\
\witem{group of stereoisomers}{Q59199015} & 2nd \\
\witem{group or class of chemical substances}{Q17339814} & 2nd \\
\witem{drug class}{Q2585617} & 2nd \\
\witem{pharmacological class of \ldots}{Q55499636} & 2nd \\
\witem{group or class of proteins}{Q84467700} & 2nd \\
\witem{structural class of chemical entities}{Q47154513} & 2nd \\
\witem{ homologous series}{Q741844} & 2nd \\
\witem{aircraft family}{Q15056993} & 3rd \\
\witem{combat aircraft family}{Q124054738} & 3rd \\
\witem{combat vehicle family}{Q100709275} & 3rd \\
\end{tabular}
\begin{tabular}{ll}
\witem{torpedo family}{Q124054628} & 3rd \\
\witem{firearm family}{Q124054676} & 3rd \\
\witem{artillery family}{Q124056256} & 3rd \\
\witem{weapon family}{Q15142889} & 3rd \\
\witem{missile family}{Q18487018} & 3rd \\
\witem{ammunition family}{Q42314051} & 3rd \\
\witem{group or class of strains}{Q75913269} & 2nd \\
\witem{film rendition of song}{Q124354051} & 2nd \\
\witem{prosodic group}{Q120733671} & 2nd  \\
\witem{metaclass of ambassadors}{Q29918297} & 3rd \\
\witem{type of fixed-size set}{Q99469810} & 3rd \\
\witem{taxonomic rank}{Q427626} & 3rd \\
\witem{medical metaclass}{Q103994247} & 3rd \\
\witem{anatomical metaclass}{Q94945104} & 3rd \\
\witem{direct anatomical metaclass}{Q103997018} & 3rd \\
\witem{indirect anatomical metaclass}{Q103997133} & 3rd \\
\end{tabular}
} \\
\noindent
Also change \witem{vehicle family}{Q22999537} from second-order class to third-order class.

\noindent
{\em Overlap between third-order classes and second-order classes after changes, and determined class order:}

\noindent
{\small
\begin{tabular}{ll}
\witem{headphones model series}{Q113711211} & 3rd\\ 
\witem{earbuds model series}{Q113711189} & 3rd\\ 
\witem{outboard motor model series}{Q104236176} & 3rd\\ 
\witem{integrated circuit series}{Q120451300} & 3rd\\ 
\witem{Sonalika DI tractors}{Q124571039} & 3rd\\ 
\witem{Navire de Sauvetage Côtier 3}{Q98960455} & 3rd\\ 
\witem{spacecraft series}{Q113255208} & 3rd \\ 
\witem{spacecraft constellation}{Q967145} & 3rd\\ 
\witem{satellite internet \ldots}{Q54806917} & 3rd\\ 
\witem{Multifunctional Transport \ldots}{Q4390199} & 3rd\\ 
\witem{Iridium NEXT}{Q3154356} & 3rd\\ 
\witem{Advanced Extremely \ldots}{Q379544} & 3rd\\ 
\witem{PROBA}{Q3888188} & 3rd\\ 
\witem{HP Pavilion}{Q3125178} & 3rd\\ 
\witem{}{Q31890935}& 3rd\\ 
\witem{rocket class}{Q114570820} & 3rd\\ 
\witem{pathovar}{Q1972414} & 2nd\\ 
\witem{set of anatomical entities type}{Q113546394} & 2nd\\ 
\witem{mineral subclass}{Q3965272} & 2nd\\ 
\end{tabular}
}

\noindent
Several of these classes were incorrectly subclasses of second-order classes, which were fixed.

\subsubsection{Instance Loops} ~ \\

\noindent
Fixing instance loops also produced significant changes so both queries and results are given here.

\noindent
{\em Instance loops:}

\begin{verbatim}
SELECT DISTINCT ?c WHERE { 
  { ?c wdt:P31/wdt:P279+ ?s .
    FILTER(?c = ?s) }
  UNION { ?c wdt:P31 ?c }
}
\end{verbatim}

\begin{figure*}[t]
\begin{verbatim}
#!/bin/python
counts = dict()
names = dict()
items = dict()
file = open('results/wikidata-split.csvs', 'r')

for line in file:
    fields = line.strip().split(',http://www.wikidata.org')
    first = fields[0].split(',', 1)
    second = 'http://www.wikidata.org' + fields[1]
    second = second.split(',', 1)
    names[second[0]] = second[1]
    counts[second[0]] = counts[second[0]] + 1 if second[0] in counts else 1
    itemid = first[0].split('/')[-1]
    if second[0] in items:
        items[second[0]].append([ itemid, first[1] ])  
    else:
        items[second[0]] = [ [itemid, first[1]] ]

scounts = {k: v for k, v in sorted(counts.items(), key=lambda item: item[1])}
running = 0
for k,v in scounts.items():
    running = running + v
    id = k.split('/')[-1]
    print(f"{running}	{v}	{id}	{names[k]}	{items[k][0:9]}")
\end{verbatim}
\caption{Items per split-order class}
\label{split program}
\end{figure*}

\noindent
{\em Instance loops results and status:}

\noindent
{\small
\begin{tabular}{ll}
\witemr{Q23958852}{variable-order class} & OK \\
\witemr{Q35120}{entity} & OK \\
\witemr{Q7048977}{abstract entity} & OK \\
\witemr{Q99527517}{collective entity} & OK \\
\witemr{Q16889133}{class} & OK \\
\witemr{Q4479242}{class} & OK \\
\witemr{Q5127848}{class} & OK \\
\witemr{Q105337053}{Capitán de amigos}& bad \\
\witemr{Q11083290}{Tangible Cultural Heritage of \ldots}& bad \\
\witemr{Q111379221}{Tcharabaou}& bad \\
\witemr{Q111441946}{Achinakrom Senior High School Wassce}& bad \\
\end{tabular}
\begin{tabular}{ll}
\witemr{Q115604369}{Mosque in Ngawi City}& bad \\
\witemr{Q115777074}{}& bad \\
\witemr{Q115793804}{}& bad \\
\witemr{Q115796374}{}& bad \\
\witemr{Q115827918}{Gedung Depo Arsip Kota Sukabumi}& bad \\
\witemr{Q115907462}{}& bad \\
\witemr{Q116739520}{}& bad \\
\witemr{Q116884870}{}& bad \\
\witemr{Q116907856}{Zemský věstník pro Čechy}& bad \\
\witemr{Q117012228}{Carmelle Pilon Studio}& bad \\
\witemr{Q11702283}{Real Club de Tenis de San Sebastián}& bad \\
\witemr{Q117325044}{Djeregou}& bad \\
\witemr{Q117325589}{}& bad \\
\witemr{Q117326036}{}& bad \\
\witemr{Q117847876}{}& bad \\
\witemr{Q118225379}{}& bad \\
\witemr{Q118669954}{Bearings}& bad \\
\witemr{Q118756027}{Nysc secretariate}& bad \\
\witemr{Q119179672}{}& bad \\
\witemr{Q119574160}{Goddy O. Onyekaonwu}& bad \\
\witemr{Q119585446}{Tectonic maps}& bad \\
\witemr{Q119854469}{Hotel San Giorgio}& bad \\
\witemr{Q119891003}{Utilization of nutritional compounds \ldots}& bad \\
\witemr{Q120369817}{Water scarcity in Nigeria}& bad \\
\witemr{Q120735155}{}& bad \\
\witemr{Q120829234}{Sea channel of the Gulf of Ob}& bad \\
\witemr{Q121464020}{Diagnostic Laboratory Services, Inc.}& bad \\
\witemr{Q121535142}{Binuang Kingdom}& bad \\
\witemr{Q121788204}{}& bad \\
\witemr{Q121788606}{}& bad \\
\witemr{Q122092563}{}& bad \\
\witemr{Q122270828}{}& bad \\
\witemr{Q122879332}{}& bad \\
\witemr{Q122979154}{}& bad \\
\witemr{Q123192102}{Cultural observatory}& bad \\
\witemr{Q123427080}{Erazik Grigoryan}& bad \\
\witemr{Q123563283}{\ldots}& bad \\
\witemr{Q123573904}{\ldots}& bad \\
\witemr{Q123627964}{ANNAJM LEARNERS' ACADEMY}& bad \\
\witemr{Q123736830}{Tourist Centres in Ghana}& bad \\
\witemr{Q124364553}{Wiki Loves Marche 2023 award \ldots}& bad \\
\witemr{Q124461446}{Giornate FAI}& bad \\
\witemr{Q124635068}{album malakias}& bad \\
\witemr{Q124653470}{}& bad \\
\witemr{Q124658147}{}& bad \\
\witemr{Q124663585}{}& bad \\
\witemr{Q125145676}{Telangana legislature assembly building}& bad \\
\witemr{Q125415172}{}& bad \\
\witemr{Q125472263}{}& bad \\
\witemr{Q125499829}{WELLI v OKECHUKWU \ldots}& bad \\
\witemr{Q125839978}{Gobarau Microfinance Bank Ltd}& bad \\
\witemr{Q125908142}{}& bad \\
\witemr{Q125997233}{}& bad \\
\witemr{Q126205647}{CHRISTIAN SERVICE UNIVERSITY}& bad \\
\witemr{Q126371076}{}& bad \\
\witemr{Q16670127}{battery}& bad \\
\witemr{Q16718360}{a set of minerals}& bad \\
\witemr{Q19516072}{GP Internacional do Guadiana}& bad \\
\witemr{Q20864827}{Romani cuisine}& bad \\
\witemr{Q2202418}{}& bad \\
\witemr{Q23244}{Zählsprengel}& bad \\
\witemr{Q264975}{Yade}& bad \\
\witemr{Q29161218}{Voile de la Vierge}& bad \\
\witemr{Q2986441}{ancient Egyptian funerary texts}& bad \\
\witemr{Q3121589}{Roman defensive walls of Tarragona}& bad \\
\end{tabular}
\begin{tabular}{ll}
\witemr{Q3509331}{fauna of Asia}& bad \\
\witemr{Q4114379}{}& bad \\
\witemr{Q4243465}{}& bad \\
\witemr{Q5061672}{Central Plaza 1, Brisbane}& bad \\
\witemr{Q5710018}{Astrolabe of al-Sahlî}& bad \\
\witemr{Q579036}{sounding rocket}& bad \\
\witemr{Q6124703}{}& bad \\
\witemr{Q66022902}{Didachara mosque}& bad \\
\witemr{Q6671830}{long-acting reversible contraception}& bad \\
\witemr{Q669000}{Estonian mythology}& bad \\
\witemr{Q7362325}{Roman roads in Africa}& bad \\
\witemr{Q7898928}{Upper South Province}& bad \\
\witemr{Q94440081}{Fabio Valentini}& bad \\
\witemr{Q97164204}{Phim moi}& bad \\
\witemr{Q98090405}{Maria Almasri}& bad \\
\witemr{Q98427547}{Kente Festival}& bad \\
\witemr{Q99337968}{Aishetu Dozie}& bad \\
\witemr{Q386724}{work}& bad \\
\witemr{Q114213}{Arab nationalism}& bad \\
\witemr{Q1151067}{rule}& bad \\
\witemr{Q1153563}{dead man's switch}& bad \\
\witemr{Q11658}{transformer}& bad \\
\witemr{Q116763588}{ATFX CopyTrade}& bad \\
\witemr{Q124049657}{marketing material}& bad \\
\witemr{Q1373696}{mace}& bad \\
\witemr{Q1414937}{combatant}& bad \\
\witemr{Q15096222}{Förderpreis zum Literaturpreis der \ldots}& bad \\
\witemr{Q15304504}{direct transmission}& bad \\
\witemr{Q172175}{Vajrayana}& bad \\
\witemr{Q209572}{military service}& bad \\
\witemr{Q2246475}{direct and indirect contact transmission}& bad \\
\witemr{Q22906682}{male as norm}& bad \\
\witemr{Q3450985}{neurodevelopmental disorder}& bad \\
\witemr{Q3961002}{division sign}& bad \\
\witemr{Q4110717}{types of prayer in Islam}& bad \\
\witemr{Q4254955}{horizontal transmission}& bad \\
\end{tabular}
\begin{tabular}{ll}
\witemr{Q4364514}{lamella (materials)}& bad \\
\witemr{Q4461019}{}& bad \\
\witemr{Q4857865}{bar}& bad \\
\witemr{Q5162734}{conscientious objection to military taxation}& bad \\
\witemr{Q545991}{artificial respiration}& bad \\
\witemr{Q6574422}{list of Iranian Arabs}& bad \\
\witemr{Q712838}{thyroid function tests}& bad \\
\witemr{Q853800}{trace fossil}& bad \\
\end{tabular}
}
\noindent
The ancestors of the incorrect instance loops were found by looking for instance chains of length $n$ until a fixed point was reached, with the following query reaching the fixed point:

\begin{verbatim}
SELECT DISTINCT ?c WHERE {
 VALUES ?x { [ids for the 113 bad loops] } }
  ?x wdt:P31/wdt:P279* ?b . 
  ?b wdt:P31/wdt:P279* ?a .
  ?a wdt:P31/wdt:P279* ?c .
}
\end{verbatim}

After eliminating the first group of incorrect instance loops the minimum order counts were as follows:

\noindent
\begin{tabular}{@{}c@{}r@{\hspace{0.25em}}r@{\hspace{0.25em}}r@{\hspace{0.25em}}r@{}}
\hline
Minimum & \multicolumn{2}{c}{Full} & \multicolumn{2}{c}{Instance-Only}  \\
Order & \multicolumn{1}{c}{Initial} & \multicolumn{1}{c}{Revised} & \multicolumn{1}{c}{Initial} & \multicolumn{1}{c}{Revised}  \\
\hline
1 & 19,299,681 & 19,025,409 & 115,360 & 115,467 \\
2 &     44,716 &  44662 & 13,847 & 13,661 \\
3 &      7,678 &   7,552 & 3,647 & 3,346 \\ 
4 & 	 2,854 &   1,924 & 2430 & 1,411 \\ 
5 &	 1,657 &   1,065 & 1,444 & 794 \\
6 &	 1,413 &   722 & 1,345 & 591 \\
\hline
\end{tabular}

The incorrect double instance loops were eliminated by removing \witem{instance of}{P31} links from the following classes:
\witemr{Q108782497}{Ignyte Awards},
\witemr{Q126205248}{Conjunto Hidráulico del Ruedo Y La  Laguna},
\witemr{Q2371236}{Norwegian People's Aid},
\witemr{Q111169536}{Digital sequence information},
\witemr{Q6517445}{legal abuse},
\witemr{Q12786121}{warfare},
\witemr{Q1318176}{Holy Sepulchre chapel in Głogówek},
\witemr{Q30228510}{},
\witemr{Q1548436}{European Short Track Speed Skating Championships},
\witemr{Q859170}{color scheme},
\witemr{Q4135220}{genderfuck},
\witemr{Q22952981}{Serverless Framework},
\witemr{Q5589282}{Government of Omaha},
\witemr{Q7089412}{Omaha City Council},
\witemr{Q125546900}{}, and
\witemr{Q125546978}{}.

\subsubsection{Split-Order Classes} ~ \\

The count of items per split-order class was computed by the Python program in
Figure~\ref{split program} on the split-order classes computed above.

\subsubsection{Large Number of Classes} ~ \\

\noindent
{\em Count by direct subclass of \witems{class}{Q16889133}:}

\begin{verbatim}
SELECT DISTINCT (COUNT (DISTINCT ?f) AS
     ?count) ?sub WHERE {
 { SELECT DISTINCT ?sub ?f WHERE {
  ?sub wdt:P279 wd:Q16889133 .
  ?f wdt:P31/wdt:P279* ?sub .
 } }
} GROUP BY ?sub ?subLabel
\end{verbatim}

\end{document}